\documentclass[10pt,twocolumn]{article}

\usepackage[letterpaper,margin=0.75in]{geometry}

\usepackage{microtype}
\usepackage{graphicx}
\usepackage{subcaption}
\usepackage{enumitem}
\usepackage{listings}
\usepackage{xcolor}
\usepackage[hidelinks]{hyperref}

\usepackage{amsfonts}

\newcommand{\Figure}[1]{Figure~\ref{#1}}
\newcommand{\Section}[1]{Section~\ref{#1}}
\newcommand{\Listing}[1]{Listing~\ref{#1}}
\newcommand{\Table}[1]{Table~\ref{#1}}
\definecolor{keywordblue}{RGB}{0,102,204}
\definecolor{commentgray}{RGB}{106,115,125}
\definecolor{stringred}{RGB}{163,21,21}
\definecolor{backgroundlight}{RGB}{250,250,250}
\lstdefinelanguage{MLIR}{
    morekeywords={
        scf, for, if, while, yield, func, return, module,
        arith, addf, mulf, subf, divf, addi, muli, subi, divi,
        mtia, read_vec, write_vec,
        mtiaarith, addf_vx,
        affine, memref, tensor, linalg,tl,load, arange, sum, dma_in, dma_out, range, broadcast, extract_slice,get_pe_id,get_pe_grid_size,re_reduce, make_block_ptr, make_tensor_descriptor,
        mask, other
    },
    morekeywords=[2]{vector, i32, i64, f32, f64, index},
    sensitive=true,
    morecomment=[l]{//},
    morecomment=[l]{\#},
    morestring=[b]",
}
\lstdefinestyle{mlirstyle}{
    language=MLIR,
    basicstyle=\ttfamily\scriptsize,  
    keywordstyle=\color{keywordblue}\bfseries,
    keywordstyle=[2]\color{stringred}\bfseries,
    commentstyle=\color{commentgray}\itshape,
    stringstyle=\color{stringred},
    backgroundcolor=\color{backgroundlight},
    frame=single,
    framesep=10pt,
    rulecolor=\color{black!20},
    breaklines=true,
    breakatwhitespace=false,
    xleftmargin=10pt,
    framexleftmargin=5pt,
    tabsize=2,
    showstringspaces=false,
    keepspaces=true,
    columns=fullflexible,
}

\definecolor{codegreen}{rgb}{0,0.6,0}
\definecolor{codegray}{rgb}{0.5,0.5,0.5}
\definecolor{codepurple}{rgb}{0.58,0,0.82}
\lstdefinestyle{pythonstyle}{
    language=Python,
    backgroundcolor=\color{backgroundlight},
    commentstyle=\color{codegreen}\itshape,
    keywordstyle=\color{blue}\bfseries,
    numberstyle=\tiny\color{codegray},
    stringstyle=\color{codepurple},
    basicstyle=\ttfamily\scriptsize,
    breakatwhitespace=false,
    breaklines=true,
    keepspaces=true,
    showstringspaces=false,
    tabsize=4
}

\def\BibTeX{{\rm B\kern-.05em{\sc i\kern-.025em b}\kern-.08em
    T\kern-.1667em\lower.7ex\hbox{E}\kern-.125emX}}
\begin{document}

\pdfpagewidth=8.5in
\pdfpageheight=11in

\pagenumbering{arabic}

\title{Triton for MTIA: Bridging the Programming Model Gaps for Custom AI Accelerators}
\author{%
\parbox{0.98\textwidth}{\centering
\small
Haishan Zhu, Domi Yan, Michael Levesque-Dion, Changxu Zhang, Mitch Gamburg, Kirsten Lee, Giancarlo Colmenares, Aditya Bhagwat, Arnab De, Markus Le Roux, Victor Perez Carrasco, Xin Tong, Will Cromar, Simran Barnwal, Andrew Uderian, Sridhar Gopinath, Jan Szczepaniec, Daniel Neilson, Blaine Burton Rister, Jordan Fix, Jazlyn Li, Zejun Huang, Lite Ye, Nan Zhang, Xinchen Guo, Andiry Xu, Michael Roberts, Kunming Ho, Site Cao, Suryadev Sahadevan Rajesh, Tristan Trouwen, Mike Tsai, Jake Lee, Wayne Su, Yuhan Chen, Xiaolong Xie, David Eklov, Aaron Barnes, Max Bremer, Adam Belay, Shintaro Iwasaki, Roman Levenstein, and Ajit Mathews\\[0.6em]
\textit{Meta Platforms}\\[0.4em]
}}
\date{}

\maketitle
\thispagestyle{plain}
\pagestyle{plain}

\begin{abstract}
The rapid growth in machine learning workloads has fueled the proliferation of custom accelerator architectures. Designed from the ground up, these accelerators often expose programming models that are distinct from GPUs. While hyperscalers and AI chip startups continue to innovate in this space, achieving broad operator coverage to support diverse models remains a major challenge. Additionally, an easy-to-use, high-level kernel programming language is important for rapid iteration of models and kernels. Triton, together with TorchInductor, addresses these issues on GPUs, but its viability on accelerators with different programming models has yet to be established. In this work, we present the first production-scale application of Triton on a custom ML accelerator, MTIA-2i, developed by Meta. To support MTIA-2i, we develop a new compiler backend that targets it, introduce enhancements to TorchInductor code generation, and propose minimal language extensions that expose MTIA-specific architectural features. We demonstrate that Triton-MTIA kernels achieve performance competitive with expert-tuned C++ implementations. Leveraging these development efficiency gains, we successfully deployed manually written and Inductor-generated Triton kernels in production across approximately 60 different model types, accounting for 50\% of layers and 47\% of non-GEMM execution time for these models. Our results provide compelling evidence that DSLs like Triton can bridge the programming model gaps between ML frameworks, kernels, and custom accelerators, enabling rapid innovation and efficient deployment at scale.

\end{abstract}

\section{Introduction}
\label{sec:intro}

Modern ML applications target a growing range of use cases from LLMs to recommendation systems to video processing for self-driving cars~\cite{llama3, dlrm, dhen, hstu}. As a result, model diversity is increasing over time, placing a huge demand on the number of operators that must be supported. For example, PyTorch's core built-in kernel library (ATen) includes almost 200 operators, while the full ATen library includes more than 540 operators~\cite{coreAten}.

Meanwhile, the range of available ML accelerators is also growing, including GPUs from established vendors and custom ASICs developed by hyperscalers and numerous AI chip startups~\cite{TPUDesign2021, thinkFastTSP2020, SambaNova2021, awsTrainium, microsoftMAIA, tenstorrentBlackhole, cerebras}. For example, Meta relies on a combination of multiple generations of AMD and NVIDIA GPUs, along with in-house custom ASICs~\cite{coburn2025meta}. Custom ASICs have the benefit of being more tailored to particular ML workloads, but they also require developers to adopt new programming models.

The increasing diversity of ML models and hardware types poses a significant challenge for developers. Delivering good performance requires deep knowledge of the low-level details of the underlying hardware (memory management, command scheduling, etc.). As a result, kernels must be hand-optimized for each type of accelerator, and these optimizations don't translate well across hardware generations. This conflicts with the need to rapidly iterate to explore new models and ways of optimizing them.

One direction that can ease this burden is to use a Domain-Specific Language (DSL) that abstracts away the complexity of supporting many models and different types of hardware. A popular example of this is Triton~\cite{triton}, which is a specialized language and compiler (integrated with PyTorch) for writing high-performance kernels. Triton is gaining traction in the industry because it hides many low-level hardware details and allows developers to write code that is portable across multiple accelerators. However, Triton was initially designed for, and currently only supports, accelerators with GPU-style programming models (e.g., NVIDIA and AMD), and it has not yet been demonstrated that it can be used with other types of accelerators.

In this paper, we ask the following question: \textit{can we support the full range of ML accelerators that are deployed in today's datacenters without sacrificing developer productivity?} To help answer this question, we study whether we can make Triton compatible with \emph{MTIA-2i}~\cite{coburn2025meta, mtia2023} at production scale. MTIA-2i is a second-generation custom ASIC designed at Meta that targets ML inference workloads. We believe MTIA-2i provides a useful data point because it has a hardware programming model that differs significantly from GPUs.

These differences manifest in a few ways. First, on MTIA-2i, kernels issue commands asynchronously to fixed-function units that operate on blocks of data, rather than instructions in threads working on individual elements like on GPUs. Second, MTIA-2i programmers are also responsible for managing on-chip memory through a FIFO abstraction, instead of more automatically through a hardware-managed cache hierarchy like on GPUs. Finally, careful command scheduling is required on MTIA-2i to pipeline operations and maximize throughput, instead of relying on hardware thread schedulers to hide instruction latency like on GPUs.

To overcome these challenges, we developed a new Triton compiler backend for MTIA-2i. This compiler can lower Triton kernels that were originally authored for GPUs and generate code for MTIA-2i, while applying a set of performance optimization passes during the compilation process. We further enabled the TorchInductor-Triton flow by tailoring TorchInductor code generation to account for architectural constraints and expanding the use of Triton templates, thereby improving both functionality coverage and performance. Finally, for cases where even higher performance is needed, we developed several language extensions that target MTIA-2i specifically. This allows kernel authors to have direct control over kernel behavior that would otherwise be difficult for the compiler to infer.

We have successfully deployed both manual and Inductor-generated Triton kernels in production. Within the period of a quarter, we tripled the number of model types that employ Triton and doubled the runtime spent in Triton on MTIA-2i, while providing performance and accuracy comparable to those of expert-tuned kernels authored using low-level APIs. This achievement highlights the impact of higher-level DSLs and ML framework integration on development velocity for custom accelerators.

In summary, our main contributions are as follows:
\begin{itemize}[noitemsep, nolistsep]
\item We demonstrate that DSLs with high-level syntax (e.g., Triton) can be used to author efficient kernels not only for GPUs but also on ML accelerators with highly customized programming models, and compilers can successfully generate code that leverages ML accelerator features to reach competitive performance. 
\item We present minimal language extensions and related compiler-kernel co-optimizations that unlock the Triton language for authoring highly optimized kernels on MTIA, and provide an in-depth study comparing the generated code performance with expert-tuned low-level C++ code.
\item We discuss our experience deploying both manual and Inductor-generated Triton kernels for MTIA in production, accounting for 50\% of layers and 47\% of non-GEMM execution time for 60 model types. In particular, we tripled the number of model types using Triton and doubled its runtime over the course of a quarter. This demonstrates the benefits of improved kernel development velocity and ML framework integration.

\end{itemize}

\section{Background}
\label{sec:background}

\subsection{DSLs for GPU Programming}

Several DSLs have been proposed to improve developer productivity in developing high-performance kernels for GPUs. Triton is one of the most popular options~\cite{triton}. It was originally developed to simplify GPU programming by abstracting away low-level hardware details and enabling developers to write efficient kernels using a Python-like syntax. More recently, Triton has become the de facto language for manual fusion and optimization at the kernel level. Notably, FlashAttention-V2~\cite{flashattention2} is one of the many examples~\cite{flaggems2024, pytorch2024flexattention,gptoss} where complex kernels and libraries were rapidly developed using Triton.

Operations in Triton operate on multidimensional tensors. Kernels explicitly describe loads and stores. Dependencies between operations are naturally expressed through value semantics. \Listing{list:triton-example} shows a snippet of a simple Triton kernel~\cite{triton}.

\begin{lstlisting} [
    style=pythonstyle,
    caption=A snippet of a Triton kernel,
    breaklines=true,
    frame=topline,
    label={list:triton-example},
    float
]
# Calculate address offsets of the data to work on.
block_start = pid * BLOCK_SIZE
offsets = block_start + tl.arange(0, BLOCK_SIZE)
# Create a mask to guard out-of-bounds accesses.
mask = offsets < n_elements
# Operations are performed at BLOCK_SIZE granularity
x = tl.load(x_ptr + offsets, mask=mask)
y = tl.load(y_ptr + offsets, mask=mask)
output = x + y
tl.store(output_ptr + offsets, output, mask=mask)
\end{lstlisting}

The kernel DSL landscape is advancing quickly. We briefly discuss a few examples beyond Triton: Helion~\cite{helion} is a higher-level DSL in PyTorch that further reduces manual effort and provides more automatic tuning than Triton. Pallas~\cite{jax-pallas} is an extension to JAX that provides a combination of high-level and low-level programming interfaces. TLX~\cite{tlx} is a set of low-level Triton language extensions that target GPUs. Gluon~\cite{gluon} and CuTeDSL~\cite{cutedsl} are lower-level DSLs that expose more GPU architecture details in exchange for a higher performance ceiling. Other notable examples include TileLang~\cite{wang2025tilelang}, CuTile~\cite{cuTile_SciPy2025}, and NKI~\cite{aws-neuron-jax-2024}.

\subsection{PyTorch 2.0 and Inductor}
TorchInductor, or Inductor for short, is a compiler that is part of the
PyTorch ML framework \cite{pytorch2}. Inductor takes in a traced PyTorch model and generates
optimized kernels for the operators in the model. It further automatically
fuses operators to minimize global memory transfers, specializing them to the
tensor shapes, data types, and constant values used in the model. Inductor supports several
backends, such as Triton, C++, and Halide~\cite{halide}. 

\subsection{MTIA Architectural Overview}
\label{sec:mtia-overview}

\begin{figure}[t]
\centering
\includegraphics[width=\columnwidth]{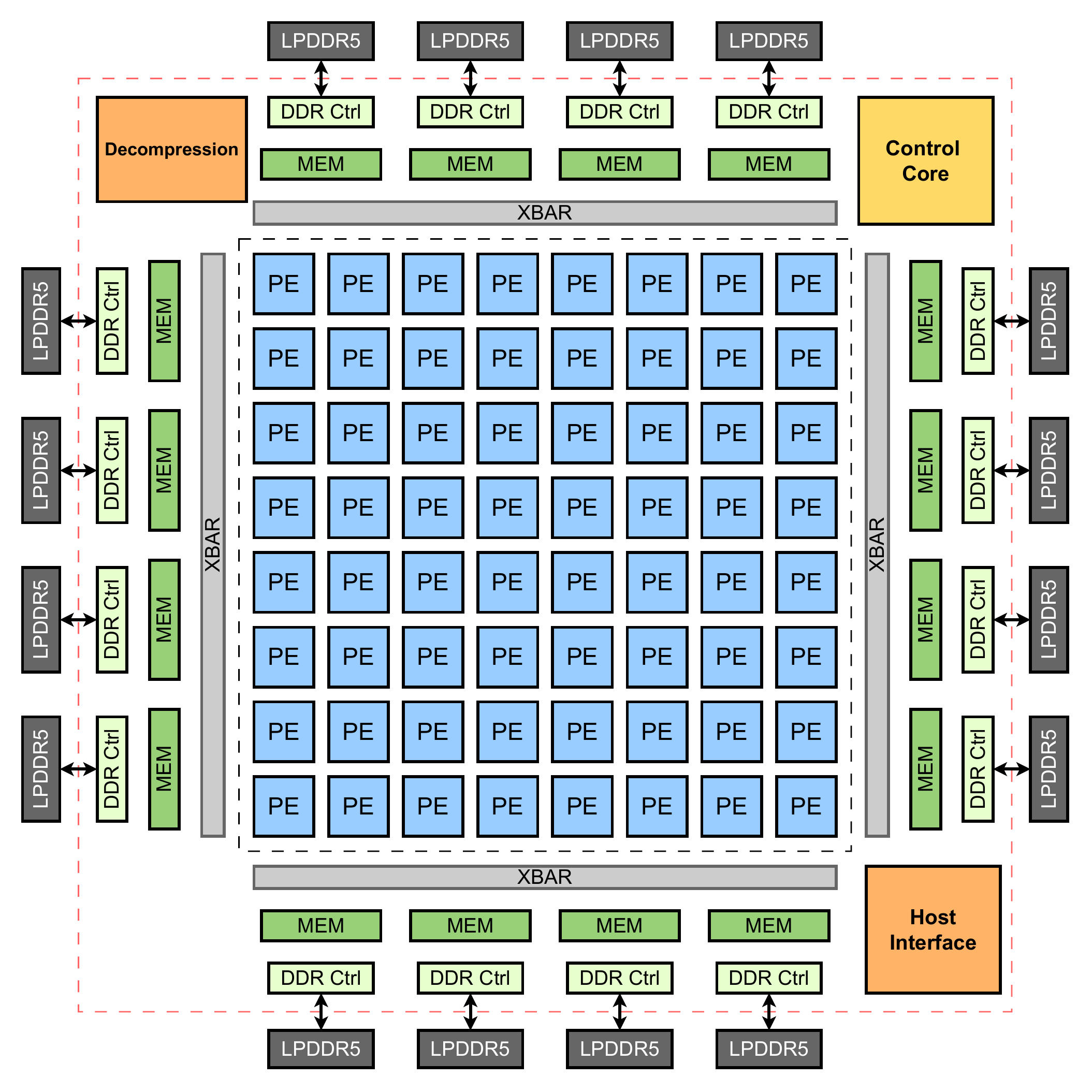}
\caption{High-level architecture of MTIA 2i.}
\label{fig:mtia-arch}
\end{figure}

\Figure{fig:mtia-arch} depicts the overall architecture of MTIA-2i~\cite{coburn2025meta, mtia2023}, which consists of an 8x8 array of \textbf{processing elements (PE)} connected via a
customized network-on-chip (NoC). The grid connects to a set of on-chip
memory blocks and off-chip memory controllers through crossbars on each side.

\begin{figure}[t]
\centering
\includegraphics[width=0.9\columnwidth]{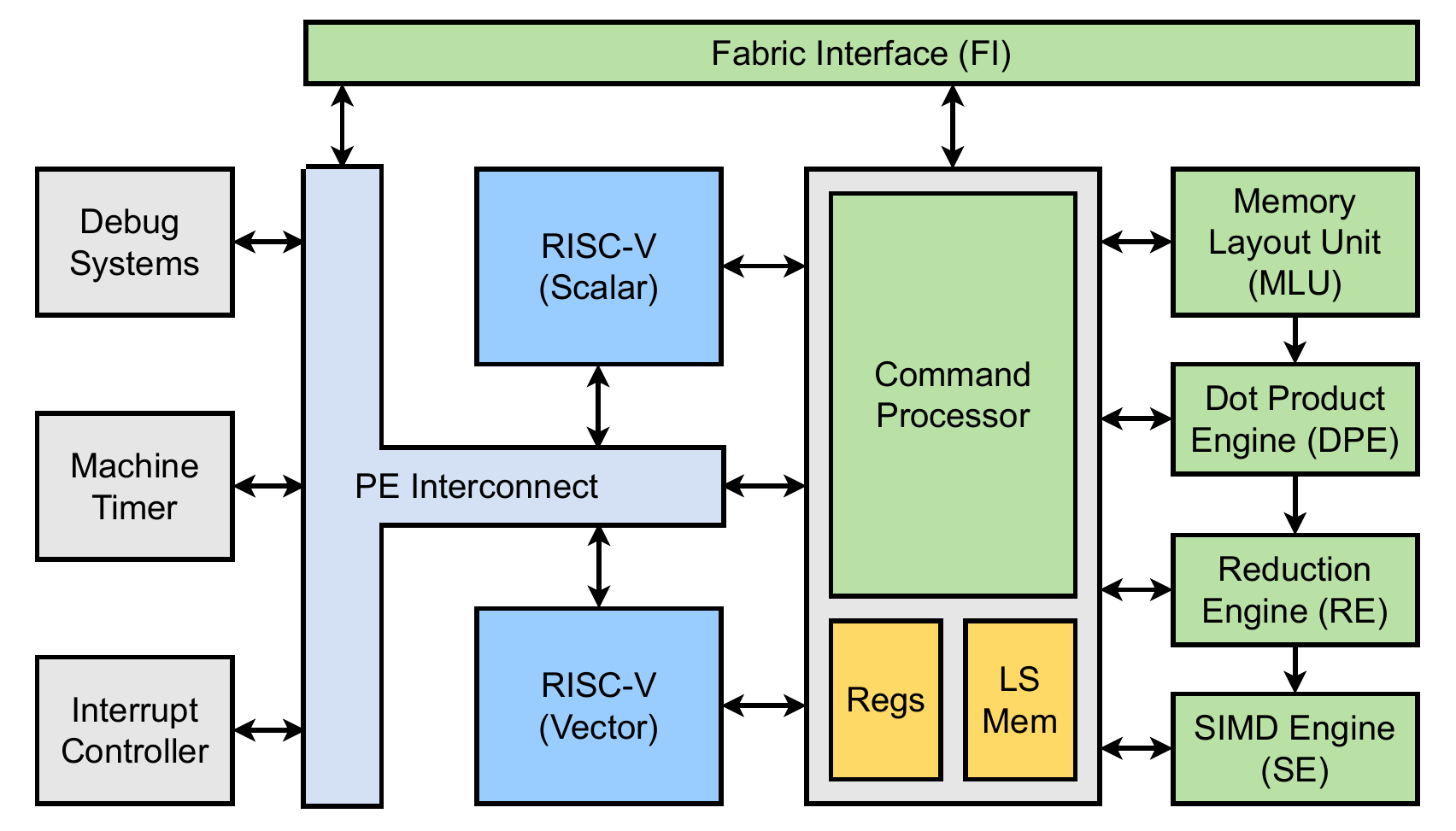}
\caption{PE's internal organization.}
\label{fig:pe-arch}
\end{figure}

\Figure{fig:pe-arch} shows the internal architecture of a PE. Each PE comprises two RISC-V processor cores and the following \textbf{Fixed-Function Units (FFUs)}.

\begin{itemize}
    \item \textbf{Command Processor (CP)}: Manages execution, dependency
        checking, and scheduling for FFUs.
    \item \textbf{Dot Product Engine (DPE)}: Performs General Matrix
        Multiplication (GEMM) operations.
    \item \textbf{SIMD Engine (SE)}: Performs high-performance vector operations:
        quantization, nonlinearities, and reductions.
    \item \textbf{Reduction Engine (RE)}: Accumulates DPE
        results; forwards results to neighboring PEs or local SE.
    \item \textbf{Memory Layout Unit (MLU)}: Handles memory layout
        transformations: transpose, concatenate, and reshape.
    \item \textbf{Fabric Interface (FI)}: DMA between Local Memory, NoC and off-chip memory.
    \item \textbf{RISC-V Vector Core}: Provides comprehensive support for
        element-wise and reduction operations.
\end{itemize}

\begin{figure}[t]
\centering
\includegraphics[width=\columnwidth]{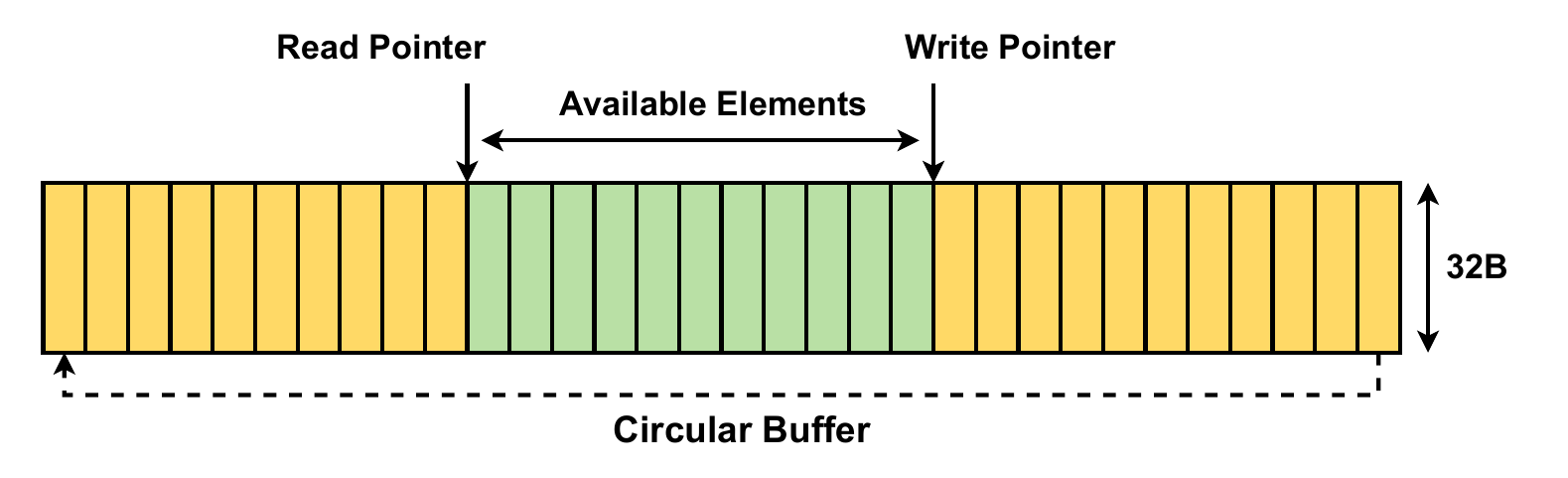}
\caption{CB abstraction overview.}
\label{fig:cb}
\end{figure}

Each PE includes a fast \textbf{Local Memory (LS)}. Notably,
data in LS is managed using a \textbf{Circular Buffer (CB)} abstraction (see
\Figure{fig:cb}). Each CB is equipped with a read pointer and
write pointer to enforce dependencies.
For instance, a read-after-write (RAW) dependency is identified by observing a
read from an address range that is past the write-pointer. In
this case, the read operation is blocked until the write-pointer has been
advanced (i.e., after a writer has filled in enough data). This abstraction
allows simultaneous reading from and writing to memory in the same CB, akin to
a FIFO that can be simultaneously enqueued and dequeued. This is the main
mechanism by which LS capacity is used; FFUs directly interact with CBs in LS. 

\subsection{Programming Models of Custom ML Accelerators}
The industry has explored a wide variety of different programming models for ML accelerators. Google’s TPUs present a VLIW model
and multiple levels of scratchpad memory backed by HBM \cite{TPUDesign2021}. Groq introduces
a Tensor Streaming Processor (TSP) architecture that exposes a stream
programming abstraction between producer and consumer on-chip engines \cite{thinkFastTSP2020}.
SambaNova introduces a Reconfigurable Dataflow Architecture (RDA) that relies
on compilers to map dataflow graphs in models onto compute units \cite{SambaNova2021}. Other
notable examples of custom accelerator architectures include Amazon Trainium
\cite{awsTrainium}, Microsoft MAIA \cite{microsoftMAIA}, Tenstorrent Blackhole \cite{tenstorrentBlackhole}, and Cerebras \cite{cerebras}.

\subsection{MLIR}
The Triton compiler for MTIA-2i is built on top of MLIR \cite{lattner2020mlir}, an open-source and extensible compiler infrastructure that has gained broad adoption within the ML compiler community. MLIR accelerates compiler development by making it easy to define custom dialects and lowering passes. It also offers a comprehensive suite of built-in dialects and compilation passes which directly enable a wide range of common transformations and optimizations. 

\section{A Triton Compiler for MTIA-{2i}}
\label{sec:compiler}

Compiling Triton for MTIA presents significant challenges due to the architectural differences between MTIA and GPUs. In this section, we highlight the major challenges and present an overview of the general compilation strategy. We then walk through the compilation process at a high level.

\subsection{Challenges and Compilation Strategy}
Providing a backend for MTIA-2i in Triton was difficult because of the following programming model differences compared to GPUs:

\begin{itemize}
\item \textbf{Execution model:} In MTIA-2i, RISC-V cores asynchronously issue commands to FFUs that operate on blocks of data. In contrast, GPUs provide a SIMT model where threads execute in lock step in a warp, with each thread processing its own data element.
\item \textbf{On-chip memory management:} In MTIA-2i, CBs provide a FIFO abstraction for managing device memory. Kernels must explicitly manage DMA transfers between main memory and CBs. In contrast, GPUs feature a hardware-managed cached memory hierarchy and a software-managed, addressable shared memory.
\item \textbf{Instruction scheduling:} In MTIA-2i, programs on RISC-V cores use software pipelining to hide computation and memory access latencies. In GPUs, a hardware scheduler hides instruction latency by context switching between warps.
\end{itemize}

These differences lead to new constraints and optimization problems which are incompatible with Triton's existing open-source compiler that targets GPUs. Consequently, we developed a new compiler for MTIA-2i from the ground up. 

\begin{figure}[t]
\centering
\includegraphics[width=0.75\columnwidth]{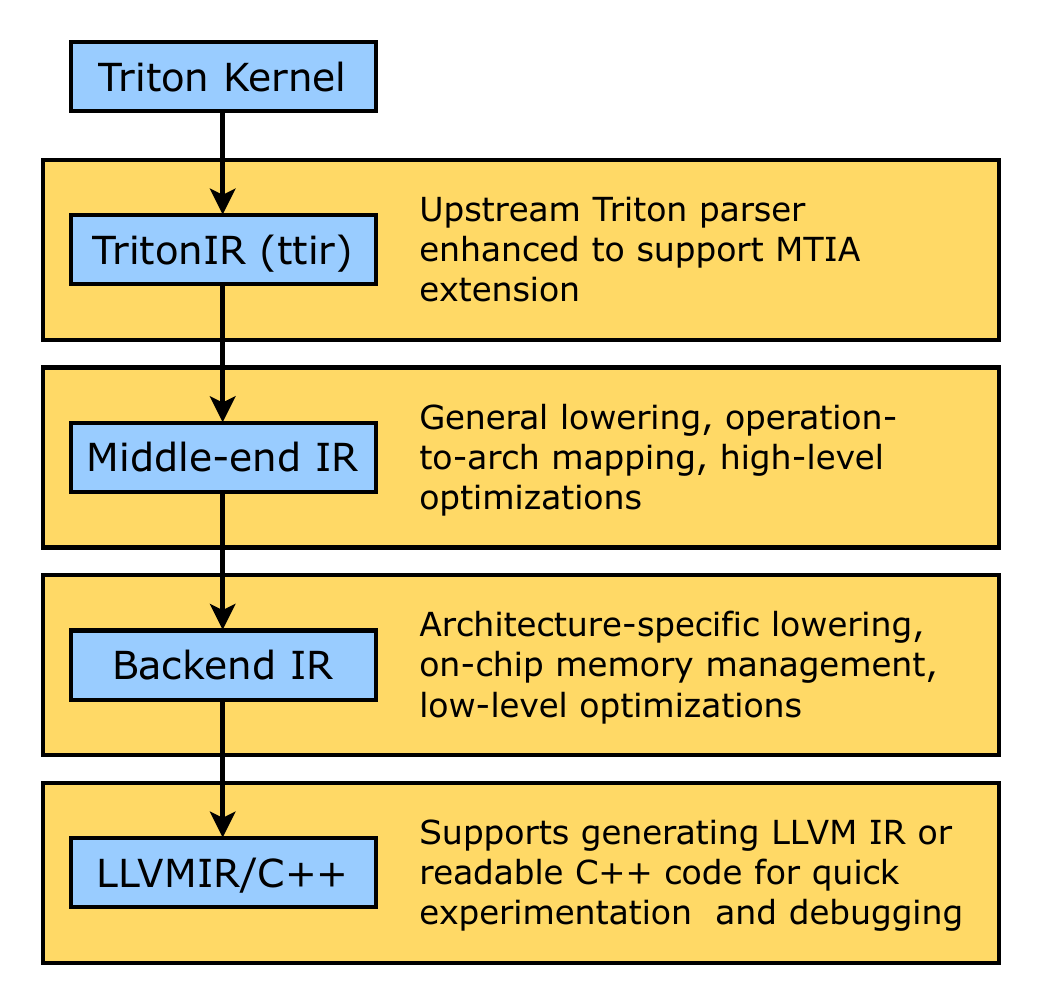}
\caption{High-level flow of Triton-MTIA Compiler.}
\label{fig:compiler}
\end{figure}

The compilation process is divided into four main stages as depicted in \Figure{fig:compiler}. The compiler leverages the upstream Triton parser as the frontend. In the middle- and backend, the compiler gradually lowers the operations by mapping them to various hardware features with increasing levels of detail. The compiler eventually generates LLVM IR or human-readable C++ code, which is passed to Clang for final compilation and linking.

At a high level, the compiler attempts to greedily leverage Fixed-Function Units (FFUs) for highly efficient block-wise computation (e.g., using the DPE, SE, MLU, RE) and data movement (using the DMA engine in FI). Tasks
that cannot be accelerated by FFUs are instead handled by the RISC-V cores. The compiler
uses the CB abstraction to manage on-chip data and to pipeline memory
accesses and compute operations. Optimizations are applied across the compilation process; we present some examples in \Section{sec:optimizations}.

\subsection{Parsing: Compatible at Language Level}
\label{sec:parsing}
We leverage the existing Triton parser to create
the initial Triton IR (ttir) from kernels. By directly building on the upstream parser
and MLIR dialects, we ensure the compiler remains up-to-date and compatible with the latest Triton language as it evolves. We further enhance the parser to support our language extensions (see~\Section{sec:extension}).

\subsection{Middle-End: Mapping Operations to Architecture}
\label{sec:middleend}
We now map operations represented in ttir onto
various hardware engines in PEs. The compiler accounts for details of
the operations, device capabilities, and performance of generated code when making these
decisions. We focus on memory and compute operations below; other miscellaneous operations
(layout transformation, synchronization, etc.) are omitted for brevity.

The main challenge for the Triton compiler in lowering memory accesses is determining whether such operations can be lowered to DMAs or must be implemented as scalar loads and stores issued by the RISC-V cores. DMAs achieve much higher memory bandwidth utilization, but
only support ``structured'' memory accesses (i.e., a multidimensional memory access that
can be expressed using a base address, dimension, and strides for each dimension).

The Triton language supports three ways of expressing memory accesses:
tensor-of-pointers, pointer-of-tensor, and tensor
descriptor. See \Listing{list:ptr-arith} for examples; note that
tensor-of-pointers (\Listing{list:ptr-arith}(a)) is flexible and can
express arbitrary scatter/gather memory accesses. However, it is also the most widely used of the three options, even for ``structured'' memory accesses.

To reliably map memory accesses to DMAs when possible, we leverage the open-source Triton IR analysis pass from \lstinline{triton-shared} \cite{triton-shared-github}. The pass analyzes pointer arithmetic operations
in the IR and rewrites them to explicitly encode structured information if possible. Unconverted memory accesses are lowered to scalar loads and stores executed by RISC-V cores.

\begin{lstlisting} [
    style=pythonstyle,
    caption=Memory accesses and middle-end lowering,
    breaklines=true,
    frame=topline,
    label={list:ptr-arith}
]
# (a) tensor of pointers
x_ptr = x_ptr + row_offsets + col_offsets
x = tl.load(x_ptr,...)

# (b) pointer of tensor
x_ptr = tl.make_block_ptr(x_ptr, shape=[M, N], ...)
x = tl.load(x_ptr)

# (c) tensor descriptor
x_desc = tl.make_tensor_descriptor(x_ptr, shape=[M, N], ...)
x = x_desc.load([0, 0])

# (d) resulting IR
%desc = tts.make_tptr %x_ptr, ...
\end{lstlisting}

For compute operations, the compiler must decide whether an operation should be executed on the FFUs (e.g., DMAs, SE, DPE, etc.) or on the RISC-V vector core. FFUs provide best performance and are designed to support the most common operators (e.g. matrix multiplications on the DPE, element-wise and reduction operations on the SE). On the other hand, the RISC-V cores support a wider range of operations and data types but at lower performance. This compilation stage makes mapping
decisions based on architecture constraints and the input IR.

\subsection{Backend: Compilation with Lower-Level Architecture Details}
\label{sec:backend}
Compared to the middle-end, the compiler backend takes additional architecture
details into consideration and further optimizes the IR. 

One of the compilation steps in this stage is tensor
bufferization. As discussed previously, the Triton language only models off-chip memory and on-chip tensors; it does not differentiate between 
different levels of the on-device memory hierarchy. The Triton-MTIA compiler
is thus responsible for allocating and reusing PE local memory. On MTIA, LS exposes a CB abstraction which implements a FIFO queue in
the architecture (see
\Section{sec:mtia-overview}). To map tensors to LS, the compiler analyzes the data usage patterns to assign memory to tensors and
manage CBs' read and write pointers.

\subsection{Code Generation: Enabling Easy Manual Examination}
The Triton-MTIA compiler supports two code generation formats: LLVM IR and C++.
After this phase, we pass the output to Clang for final
compilation and linking to create the device binary.

While LLVM IR lowering has been the standard code generation path, we have seen more uses of the C++ code generation format among MTIA expert kernel developers. In this flow,
the compiler generates human-readable C++ code that can be easily mapped back
to the original Triton source. We further enable direct
modification of the generated C++ code before final compilation and linking.
This shortens the compiler and kernel 
debugging cycle. Overall, we observe comparable performance for almost all
kernels between the LLVM IR and C++ codegen formats.

\section{Optimization Strategies in Triton MTIA Compiler}
\label{sec:optimizations}
We employ three categories of performance optimization strategies in the Triton-MTIA compiler: (1) software pipelining, (2) vector codegen optimizations, and (3) load balancing across PEs. 

\subsection{Software Pipelining}
Software pipelining is a commonly used optimization technique \cite{software-pipelining}. However,
the unique combination of the Triton language and the MTIA architecture brings many
interesting challenges. 

\subsubsection{Effective Shape Propagation}
A Triton operation operates on on-chip tensors of static shapes called block sizes. When tensor sizes are not multiples of block sizes, kernels calculate the effective shape of the tensor with masks in memory
accesses. 
We can optimize the code by extracting the effective shape and propagating this information to consumers
of the tensors. 

\begin{lstlisting} [
    style=pythonstyle,
    caption=Example of calculating dynamic tensor shape using masks in memory accesses,
    breaklines=true,
    frame=topline,
    label={list:mask-shape},
]
offset = tl.arange(0, BLOCK_SIZE)
ptr = in_ptr + offset
mask =  offset < row_size
x = tl.load(ptr, mask=mask, other=0.0)
s = tl.sum(x, axis=0)
\end{lstlisting}

\Listing{list:mask-shape} shows a kernel snippet where boundary and padding
information can be propagated. Through IR analysis, we can determine that the
tensor \lstinline{x} has a static shape of \lstinline{BLOCK_SIZE}, but computation
is only effective on \lstinline{row_size} elements.  

Gaining this insight enables a suite of optimization opportunities:
\begin{itemize}
    \item Skip initializing buffer. The analysis examines the recursive users of the loaded values to check if filling the tensor with \lstinline{other} values is necessary, and lets us skip initializing the buffer altogether if possible.
    \item Skip unnecessary computation along def-use chains. For example, in the example above, we only need to perform the sum on \lstinline{min(row_size, BLOCK_SIZE)} elements. This benefit is amplified when there is a chain of operations, or for larger or higher-dimensional tensors.
    \item Avoid materializing masks. We can skip all overheads associated with calculating the mask, including both on-chip memory usage and runtime instructions execution.  
\end{itemize}

\subsubsection{Instruction Issue Bottleneck}
As discussed in~\Section{sec:mtia-overview}, each PE is equipped with two
RISC-V cores, both of which can issue commands to FFUs. The main
goal of this design is to avoid instruction issue bottlenecks within
PEs.

In mapping operations to cores, the compiler has to take extra care in
synchronizing FFU commands that operate on CBs shared by both cores, and
accounting for the extra cycles one of the cores spends on issuing vector instructions. 

We create a general scheduling algorithm which statically analyzes the IR and
distributes commands between cores, and eventually generate code that issue FFU commands from both RISC-V cores. For even deeper kernel-level optimization, we extend the language with hints to
allow kernel authors to explicitly specify which core an operation should prefer. 
This allows expert kernel authors to make explicit command scheduling decisions. 
See~\Section{sec:extension} for more detailed examples on language extensions.

\subsubsection{Greedy CB Sizing}
MTIA architecture is designed to simplify the construction of software pipelining. In particular, RISC-V cores issue commands to FFUs asynchronously and data dependencies are enforced by participating FFUs using CB abstraction. This design allows RISC-V cores to run ahead in its execution, pre-issue as many commands to FFUs as possible, and keep the entire pipeline of memory accesses and on-chip computation occupied.

To take advantage of this design, the compiler analyzes the IR to identify the capacity requirements for all tensors in the kernel, which is made easy given that all tensors in Triton are statically sized. The compiler then greedily chooses the smallest multiplier in allocating capacity for each tensor, so that all allocations still fit in LS. Since kernels often contain an inner loop (e.g., to iterate over tensors), this optimization allows the resulting code to pipeline memory and compute operations across loop iterations.

\subsubsection{CB Pointer Management}
As discussed in~\Section{sec:mtia-overview}, LS memory is managed using a
CB abstraction. Synchronization between producers and
consumers is coordinated via explicit read and write pointer adjustments. The compiler must accurately
determine when to perform these updates to guarantee correctness.

Triton kernels may contain arbitrary control flow (e.g., loops, 
branches, nested controls, etc.). As a result, it is not always obvious where read and write pointer adjustments should be inserted.

We developed an algorithm that inserts such pointer adjustments based on how each buffer is used. The algorithm groups operations into
static ranges that read from or write to a buffer, and increments pointers at
the boundaries between ranges. This approach has proven robust across a wide
range of Triton kernels. See \Listing{list:cb-ptr-adj} for an example.

To enable software pipelining, the compiler allocates CBs whose size is a multiple of the tile size used by the kernel. Since CP issues commands to FFUs asynchronously and data dependencies are directly enforced via the CB abstraction, multiple commands can be in flight at the same time as long as there is enough space in the relevant CBs.

\begin{lstlisting} [
    caption=CB pointer adjustment insertion example,
    breaklines=true,
    frame=topline,
    label={list:cb-ptr-adj},
]

mtia.dma_in %cb, ...
// increment write pointer by %cb size
scf.for %i in (...) {
    %offset_i = ...
    %0 = mtia.read_vec %cb, %offset_i, ...
    // increment read pointer by 64
    %1 = arith.addf %0, ...
    mtia.write_vec %cb, %2, %offset_i, ...
    // increment write pointer by 64
}
mtia.dma_out %cb, ...
// increment read pointer by %cb size
\end{lstlisting}

\subsection{Utilizing RISC-V Vector Core For Long-tail Operations}
\label{sec:rvv}
In MTIA-2i, one of the RISC-V CPU cores supports the RISC-V Vector Extensions
(RVV) \cite{rvv}. We leverage this capability to accelerate ``long tail'' operations that
cannot be directly mapped to FFUs. One of the steps in RVV codegen is vectorization.
At a high-level, this process consists of two main steps:

\textbf{Chunking operations into vector sizes.} We utilize a combination of custom and
MLIR upstream vectorization passes~\cite{mlir2021} to chunk operations into the desired
vector sizes, including handling cases such as the vectorization of reduction
loops.

\textbf{Mapping MLIR Vector Operations to RVV primitives.} Most arithmetic operations
can be directly lowered to corresponding RISC-V Vector (RVV) instructions. For instance:\\
\lstinline{
can be directly mapped to the equivalent RISC-V intrinsic:\\
\lstinline{vint32m2_t vadd_vv_i32m2(vint32m2_t, vint32m2_t, size_t)}

We perform additional optimization throughout the vectorization process. We
discuss two interesting examples below.

\subsubsection{Optimizing Vector-Scalar Operations}
The upstream Triton parser always lowers tensor-scalar operations as tensor-tensor
operations with the scalar operand broadcasted to the same shape as the tensor.
Materializing the broadcasted scalar is sub-optimal because it takes both extra time and storage to compute.
As part of the vectorization process, we avoid this by lowering such operations
to vector-scalar operations defined in a custom \lstinline{mtiaarith} dialect as demonstrated in ~\Listing{list:mtiaarith}.

\begin{lstlisting} [
    caption=Vector-scalar operations optimization example,
    breaklines=true,
    frame=topline,
    label={list:mtiaarith},
    float
]
// Input IR:
%0 = vector.broadcast %arg0 : f32 to vector<64xf32>
%1 = arith.addf %arg1, %0 : vector<64xf32>

// Output IR:
%1 = mtiaarith.addf_vx %0, %arg1 : vector<64xf32>, f32 -> vector<64xf32>
\end{lstlisting}

\subsubsection{Fusing Vectorized Loops}
Intermediate results from RISC-V vector computations are stored in LS
buffers, and back-to-back vector operations require repeated round-trips
between LS and RISC-V vector registers. By fusing
the loops generated for these operations, we eliminate redundant memory
transfers and reduce instruction overhead. 

\Listing{list:fusion} demonstrates such optimization in
calculating $y = (x1 + x2)^2$ using the vector core. 

\begin{lstlisting} [
    caption=Fusion of vectorized loops example.,
    breaklines=true,
    frame=topline,
    label={list:fusion},
]
//// Before Fusion
// for loop for vector add
scf.for (...) {
 %0 = mtia.read_vec %cb0, ... : vector<64xf32>
 %1 = mtia.read_vec %cb1, ... : vector<64xf32>
 %2 = arith.addf %0, %1 : vector<64xf32>
 mtia.write_vec %cb2, %2, ... : vector<64xf32>
}
// for loop for vector multiply
scf.for (..) {
 %3 = mtia.read_vec %cb2, ... : vector<64xf32>
 # vector_multiply
 %4 = arith.mulf %3, %3: vector<64xf32>
 mtia.write_vec %cb3, %2, ... : vector<64xf32>
}

//// After Fusion
scf.for (...) {
 %0 = mtia.read_vec %cb0, ... : vector<64xf32>
 %1 = mtia.read_vec %cb1, ... : vector<64xf32>
 %2 = arith.addf %0, %1 : vector<64xf32>
 %3 = arith.mulf %2, %2: vector<64xf32>
 mtia.write_vec %cb2, %3, ... : vector<64xf32>
}
\end{lstlisting}

\subsection{Load Balancing Across PEs}
Triton PID grids can be of arbitrary size but the Processing Element (PE) grid size
on MTIA is fixed (e.g., 8 × 8 for MTIA 2i). The compiler offers a few common PID-to-PE mappings. For kernels where the physical location of PEs is
important (e.g., require synchronization or data exchange with neighboring PEs), kernel authors can also manage PID-to-PE mapping directly.

\subsubsection{Flexible PID–to-PE Mapping Strategies}
The three most common PID-to-PE mapping strategies are: linear distribution, round-robin distribution, and zigzag distribution. Since optimal workload
distribution can depend on runtime information, the compiler further provides options
for users to specify a distribution strategy in addition to the default scheme. 

\begin{figure}
\centering
\includegraphics[width=\columnwidth]{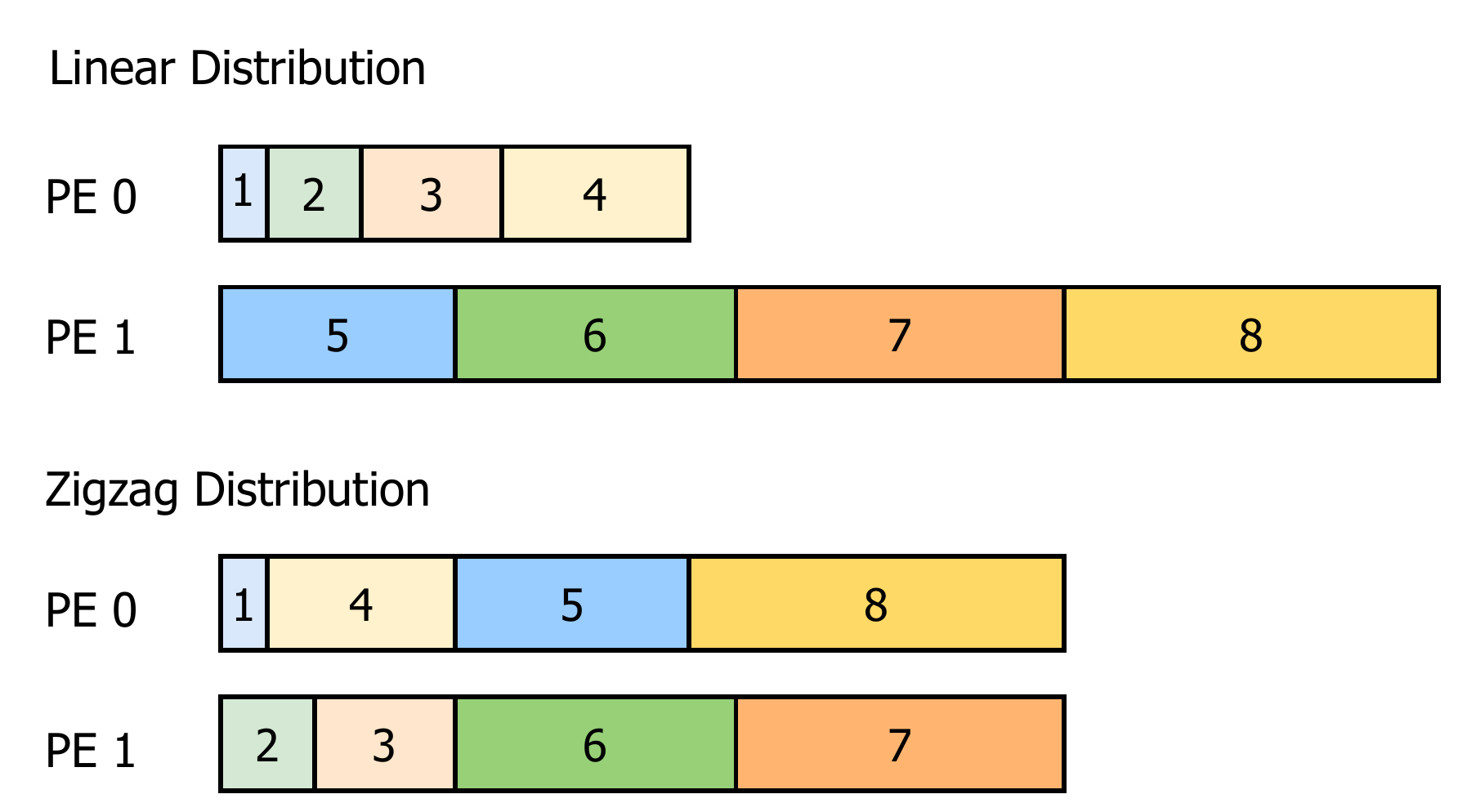}
\caption{Zigzag workload distribution compared with linear distribution. Example assumes 8 PIDs and 2 PEs; it further assumes workload increase linearly with PIDs.}
\label{fig:zigzag}
\end{figure}

To illustrate the benefit of this optimization, consider an attention
kernel \cite{attentionIsAllYouNeed} with a causal mask; each PID is responsible for one row of the
attention matrix. Using linear PID-to-PE mapping will lead to poor performance due to
uneven workload distribution (small PIDs will have less work), while a zigzag
distribution can optimally balance the workload. \Figure{fig:zigzag} depicts the application of optimization using a toy-scale example with 8 PIDs and 2 PEs.
In a production workload, we measured that the zigzag distribution provides up to an 80\% speedup over a linear distribution.

\subsubsection{User Explicit Control on Workload Distribution}
\label{sec:pid-pe-mapping}
MTIA architecture offers 2D Mesh NoC and topology-aware communication
primitives (e.g., row-wise and column-wise broadcast and reduction). To fully
utilize these features, users must explicitly control the PID-to-PE mapping.

We extended Triton to allow kernel authors to query the current
PE ID using a custom op \lstinline{tl.mtia.get_pe_id}. This approach enables
precise and efficient mapping of computation to hardware resources.

The balance between abstraction and performance is a recurring theme in this
work. \Section{sec:extension} provides more detail on language extensions.

\section{Tailoring Inductor Triton Code Generation}
MTIA’s Inductor backend shares most of its logic with the open source GPU
backends, including the core compilation passes, graph optimizations, and
Triton code generation. However, it also differs from the default GPU codegen in a
few ways. In this section, we discuss some of the interesting challenges we encountered.

\subsection{Generating Strided Memory Accesses}
While we reuse most of the Triton codegen facilities in Inductor, we pay special attention to generating memory access code. As described in \Section{sec:middleend}, MTIA
relies heavily on DMAs to guarantee performance. To ensure successful lowering to DMAs when possible, we tailor Inductor codegen to be analyzable by the Triton-MTIA compiler. 

A common case that requires such tailoring is tiling scalar indices of multidimensional tensors. For example, a program may access a 2D strided memory region with stride $s \in \mathbb{Z}^2$ and shape $d \in \mathbb{Z}^2$ using a scalar index $x \in \mathbb{Z}$ via the indexing function
$f(x)=s_1 * \lfloor x / d \rfloor +s_2*(x \bmod d)$. When we see this pattern, we tile the kernel’s iteration space to match $d$. This allows us to generate easy-to-analyze pointer arithmetic code (i.e., without modulo or division operations). Inductor can then emit \lstinline{tl.make_block_ptr} or \lstinline{tl.make_tensor_descriptor} for loads and stores, which are guaranteed to be mapped to DMAs.

\subsection{Triton Templates}
Another way we customize Inductor’s Triton codegen for MTIA is via Triton Templates, which allow us to define kernel macros that can be automatically fused with surrounding operations. While Triton Templates are commonly employed for GEMM fusion and attention kernels on GPU, on MTIA we expand their use case to other complex performance-critical kernels such as layernorm, taking advantage of architecture-optimized kernel macros. These templates are automatically fused with a variety of activation functions to eliminate loads and stores of intermediate tensors. Features of Triton templates include:
\begin{itemize}
    \item Mapping operations on N-dimensional tensors to 2D Triton kernels.
    \item Providing predictable performance on generated code by leveraging pre-optimized kernel macros.
    \item Using placeholders, mutable inputs, and boolean flags to skip unnecessary codegen and flexibly support multiple outputs and optional kernel arguments.
\end{itemize}

\subsection{Optimizing Memory and Register Use}
To make best use of the hardware, Inductor needs to model MTIA’s memory and
register resources. For example, MTIA-2i supports 15 CBs and 384 KB of LS per PE, differing considerably from the register file and local
memory available to streaming multiprocessors (SMs) on a GPU. To prevent
spilling between various levels of the memory hierarchy, MTIA’s Inductor
backend estimates the resource usage of each kernel, tuning fusion decisions and
block sizes as needed.
\section{Extending the Triton Language for MTIA}
\label{sec:extension}
Two main necessities drive our extensions to the Triton language:

\textbf{Exposing MTIA architecture features.} 
    Some MTIA features cannot be easily represented or inferred from vanilla Triton. Exposing these features at the language level allows MTIA experts
    to more effectively program the accelerator.

\textbf{Enhancing language expressiveness.} Triton is a relatively simple language by
    design. Providing more expressive operations allows kernel authors to
    better describe their intention and write less verbose code.

The rest of this section introduces several representative language extensions
that target MTIA. We focus on the intuition behind
the extensions; detailed definitions are omitted for brevity.

\subsection{Exposing MTIA Architecture Features}
\subsubsection{PE ID Querying}
As described in \Section{sec:pid-pe-mapping}, PEs on MTIA are organized in a 2D mesh
and communication between PEs is explicitly managed by software. A topology-aware
PID-to-PE mapping therefore contributes to optimal performance in
scenarios where PIDs communicate with each other (e.g., share weights, reduce
partially accumulated results, etc.) 

We introduce \lstinline{tl.mtia.get_pe_id} to allow a kernel instance to explicitly query
the PE ID to which it is mapped. Similarly, \lstinline{tl.mtia.get_pe_grid_size} queries the size of the PE grid to which all PIDs are mapped.

This seemingly simple extension provides direct utility to
kernel authors. Instead of dividing workload across PIDs which are
distributed to PEs by the compiler, kernel authors can directly manage workload-to-PE mapping
explicitly as described in~\Section{sec:pid-pe-mapping}. This also enables cross-PE
coordination as we describe next.

\subsubsection{Cross-PE Reduction}
\label{sec:cross-pe-red}
Efficient reduction operations are important for GEMM and normalization
of large tensors. On the MTIA architecture, the
Reduction Engine (RE) provides hardware acceleration by leveraging the reduction network.

We introduce \lstinline{tl.mtia.re_reduce} to allow kernels to explicitly
exercise this feature. See~\Listing{list:re-reduce} for an example. Note that
only the PE at the end of the reduction network produces the fully accumulated
result.

We also introduce a similar extension for broadcasting, another important
feature for high-performance GEMM. The details are skipped for
brevity.

\begin{lstlisting} [
    style=pythonstyle,
    caption=Semantics of tl.mtia.get\_pe\_id and tl.mtia.re\_reduce,
    breaklines=true,
    frame=topline,
    label={list:re-reduce},
]
is_top_pe = tl.mtia.get_pe_id(0) == 0
is_bottom_pe = tl.mtia.get_pe_id(0) == tl.mtia.get_pe_grid_size(0) - 1
output = tl.mtia.re_reduce("add", values, is_top_pe, is_bottom_pe)

if is_bottom_pe:
    # store result to memory
\end{lstlisting}

\subsubsection{libdevice}
Triton libdevice \cite{libdeviceAPI} is a mechanism to call library functions
defined for a specific backend. Triton-MTIA compiler takes advantage of
this API by giving users direct access to the SIMD Engine (SE), along with configurable 
compiler directives indicating if libdevice calls should lower to FFUs or to RISC-V operations.
This strategy enables seamless execution of unmodified GPU Triton kernels
on MTIA hardware while automatically applying hardware-specific optimizations, effectively
bridging the architectural differences between GPUs and MTIA without
sacrificing either portability or performance.

\subsection{Enhancing Language Expressiveness}
\subsubsection{Enabling Explicit Fine-Grained Memory Reuse}
By default, Triton does not support accessing sub-tensors; both computation and
memory operations in Triton operate on full tensors. For GPU architectures
with more elaborate hardware-managed caches, this design is perfectly reasonable
since memory regions that are repeatedly accessed will be cached. However, architectures with software-managed scratchpads (e.g., LS on MTIA) require such reuse to be explicitly expressed. 

\begin{lstlisting} [
    style=pythonstyle,
    caption=Examples of repeated data accessing using baseline Triton and explicit data reuse with tl.extract\_slice,
    breaklines=true,
    frame=topline,
    label={list:extract-slice},
]
# (a) Baseline Triton
for _ in range(...):
    # each iteration iterates through the same tensor A
    for _ in range(...):
        # each iteration uses a slice of tensor A
        a_tile = tl.load(A_ptr, ...)
        
# (b) Expressed with language extension
A = tl.load(A_ptr, ...)
for _ in range(...): 
    for _ in range(...): 
        a_tile = tl.extract_slice(A, offsets=[...])
\end{lstlisting}

\Listing{list:extract-slice}(a) provides a concrete example of repeated data access in a nested loop. Each iteration of the inner loop uses a slice of a large tensor, while each iteration of the outer loop iterate through the entire tensor. While the compiler can analyze the IR to identify reuse opportunities, such optimizations can be unreliable especially as the program becomes complicated. Kernel authors would also lose the flexibility to explicitly reuse data that has been previously loaded to LS already.

To address this gap, we introduce \lstinline{tl.extract_slice}, which allows
users to explicitly express fine-grained memory reuse patterns. Semantically,
it returns a copy of the slice of the data defined by user-specified offsets
and shape parameters; the compiler can only return a view (and avoid the extra
copy) after verifying that such a rewrite is safe. See~\Listing{list:extract-slice}(b)
for an example.

On top of the additional expressiveness, we introduce further optimizations that attempt to hide the latency of fully loading the tensor by breaking it up and sinking it into the innermost loop body. We skip details for brevity. 

\subsubsection{Custom Types and Integration In a C++-Based Software Stack}
The baseline Triton language has limited support for custom data types, which makes it difficult to share the same data structures between Triton and other parts of the MTIA software stack, which is heavily based on C++/C. For example, many existing MTIA C++ kernels work on custom data structures such as \lstinline{TensorView}s (metadata on how to address a tensor in memory) and \lstinline{CoreId} (metadata on core information). To the best of our knowledge, the most relevant work from open-source Triton is the \lstinline{@tl.core._aggregate}, which enables Triton kernels to work with Python classes. 

To close this gap, we augment Triton language with the ability to define custom types using \lstinline{ctypes.Structure}. This allows users to access fields of the custom structures, passing them by reference, even creating nested custom types. We further define utility functions to work on common custom types and data structures, in order to present a familiar programming interface to expert MTIA kernel authors. \Listing{list:custom-types} shows a brief example of how this feature can be used.

\begin{lstlisting} [
    style=pythonstyle,
    caption=Example of using Custom Types in Triton on MTIA,
    breaklines=true,
    frame=topline,
    label={list:custom-types},
    float
]
# definition of TensorView
@core.struct_type
class TensorView(ctypes.Structure):
    _fields_ = [
        ("shape", ctypes.c_uint64 * 4),
        ("stride", ctypes.c_uint64 * 4),
        ("base", ctypes.c_void_p),
        ("ndim", ctypes.c_uint32),
    ]

# Triton kernel using TensorView
@triton.jit
def ExampleKernel(
    x_view: TensorView,
):
    x_ptr = x_view.base
\end{lstlisting}

\section{Evaluation}
\label{sec:eval}

\subsection{Triton Kernel Performance}
We present performance of Triton kernels measured on MTIA-2i silicon using
configurations seen in a variety of production models. When applicable, we
compare the performance of Triton kernels to architecture roofline or
expert-tuned kernels in low-level C++. 

\subsubsection{GEMM (General Matrix Multiply)}
We evaluated GEMM kernel performance using configurations (tensor shapes,
precisions, etc.) from production recommendation and generative models.
We compare throughput metrics against the
architecture roofline.

These GEMM kernels leverage language extensions described
in~\Section{sec:extension}. We implement several variants of GEMM algorithms
with different tiling and workload distribution strategies to optimize for
different shapes and data types; a top-level dispatcher chooses the optimal variant at
runtime based on a static dispatching policy. 

As depicted in~\Figure{fig:gemm-perf}, Triton GEMM kernels consistently
achieve over 80\% of the architecture roofline and are on par with expert-tuned
kernels. 

\begin{figure}[t!]
\centering
\includegraphics[width=\columnwidth]{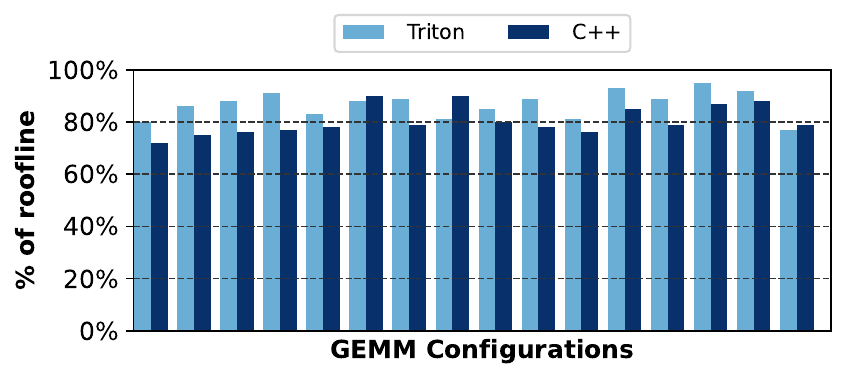}
\caption{GEMM Triton kernel performance as percentage of roofline compared with expert-tuned C++ kernels measured on MTIA silicon.}
\label{fig:gemm-perf}
\end{figure}

\begin{figure}[t!]
\centering
\begin{subfigure}{.5\columnwidth}
    \centering
    \includegraphics[width=.95\columnwidth]{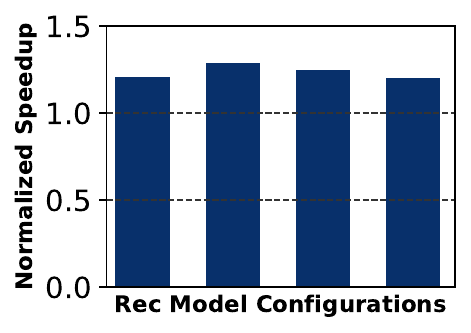}
    \caption{Recommendation Model}
    \label{fig:attention-perf-rec}
\end{subfigure}%
\begin{subfigure}{.5\columnwidth}
    \centering
    \includegraphics[width=.9\columnwidth]{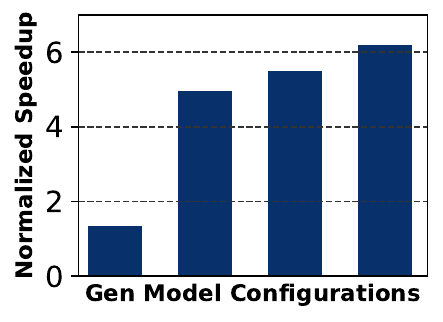}
    \caption{Generative Model}
    \label{fig:attention-perf-gen}
\end{subfigure}
\caption{Attention performance normalized to decomposed baseline measured on MTIA-2i for two model types.}
\label{fig:attention-perf}
\end{figure}

\subsubsection{Fused FlashAttention Kernels}
We evaluated and productionized a fused FlashAttention kernel on MTIA-2i silicon. 
\Figure{fig:attention-perf} illustrates its performance across configurations representative of targeted recommendation and generative models. Kernel performance is normalized to a baseline where operators are decomposed and lowered to an optimized kernel library. Notably, we observe greater speedups for generative model configurations with longer sequence lengths.

\begin{figure}[t!]
\centering
\includegraphics[width=\columnwidth]{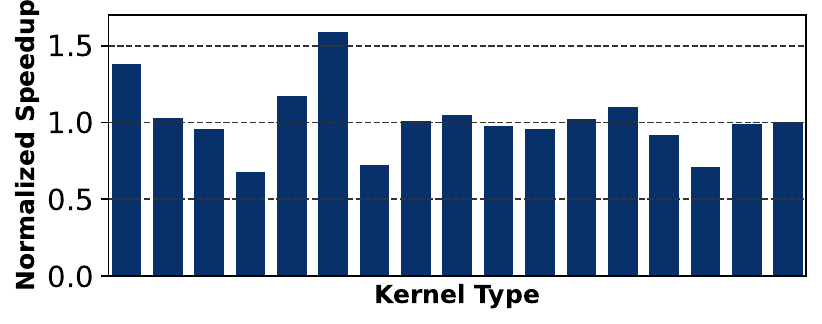}
\caption{Long tail kernel performance compared to C++ kernels.}
\label{fig:long-tail-perf}
\end{figure}

\subsubsection{Long-Tail Kernels}
We evaluated the performance of 17 long-tail kernels used in
production models on MTIA in \Figure{fig:long-tail-perf}. Each
bar plots aggregated (geomean) performance of kernels for an operator,
normalized to expert-tuned C++ kernels. In general, Triton kernels
perform on par with baseline solutions.
Notably, we are able to quickly develop these kernels and cover new scenarios and use cases for which previous kernel implementations have not been optimized (see outliers in \Figure{fig:long-tail-perf}). This
demonstrates the benefit of using Triton to accelerate kernel development
velocity.

\begin{table}[t!]
\normalsize
\caption{Operator-level coverage statistics for Inductor-based compilation flow, extracted from a set of 10 internal models.}
\label{tab:inductor-coverage-stats}
\begin{center}
\begin{tabular}{c|c|c}
    & Inference & Training \\
    \hline 
    Success & 94.5\% & 94.7\% \\
    Triton error & 4.6\% & 4.5\% \\
    Inductor error & 0.03\% & 0.01\% \\
    Misc. error & 0.5\% & 0.6\% \\ 
\end{tabular}
\end{center}
\end{table}

\subsection{Inductor-Triton Model Coverage and Performance}
To gauge the effectiveness of Inductor on MTIA, we test the coverage and performance of the generated Triton kernels on both individual operators as well as an internal model.

\subsubsection{Operator Coverage and Performance}
First we report operator-level testing. We extract detailed PyTorch operator usage information from ten internal models in both inference and
training. We then measure the pass rate of our Inductor-Triton compilation flow on PyTorch tests by replicating each unique combination of operator, data type, argument tensor shapes, and non-tensor arguments.
The coverage results are shown in \Table{tab:inductor-coverage-stats}. The
majority of the remaining failures are
attributed to limitations in the MTIA hardware (e.g., FFU constraints are incompatible with some of the test cases).

We also evaluate the performance of Inductor-generated kernels on a few representative examples. Kernel fusion is one of Inductor’s most important performance optimizations. To demonstrate the viability of
automatic fusion, we compared the performance of an automatically fused
layernorm-sigmoid kernel, which is often observed in production models, to a handwritten fused baseline. The results in \Figure{fig:inductor-fused-perf} suggest that automatic fusion is competitive with
handwritten kernels.

\begin{figure}[t!]
    \centerline{\includegraphics[width=\columnwidth]{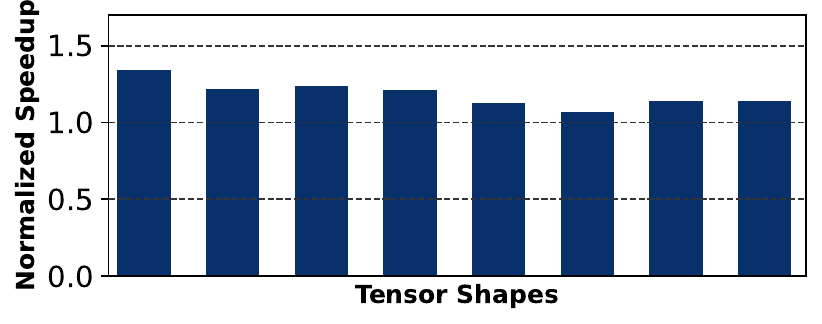}}
    \caption{Normalized speedup between a fused layernorm-sigmoid kernel generated in Inductor via Triton Templates and a handwritten baseline}
    \label{fig:inductor-fused-perf}
\end{figure}

\subsubsection{Model Coverage and Performance}

\begin{figure}[t!]
    \centerline{\includegraphics[width=\columnwidth]{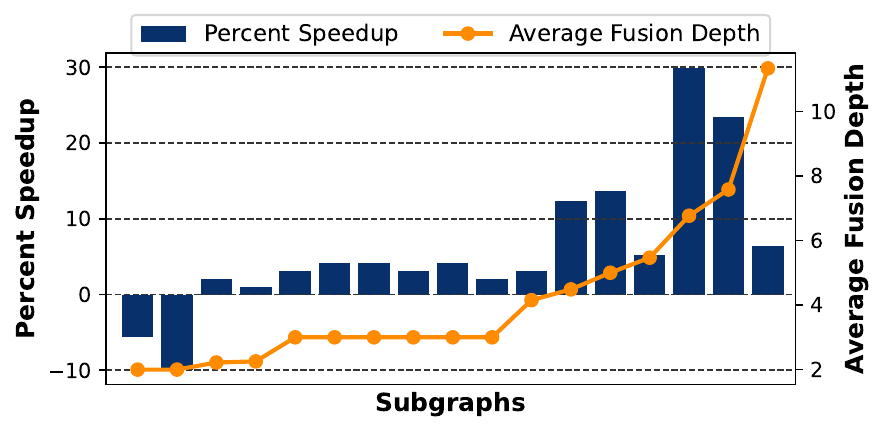}}
    \caption{Sub-graph performance of training a recommendation model. Left axis plots speedup of sub-graphs with Inductor-generated fused kernels relative to their latencies without Inductor. Right axis plots the average fusion depth in number of ATen operators.}
    \label{fig:inductor-subgraph}
\end{figure}

To understand coverage and performance of Inductor-Triton in production, we analyze the performance of training a recommendation model consisting of multiple subsections of eager- and graph-mode execution. In total, Inductor-generated kernels accounted for 22\% of graph-mode execution time on MTIA (excluding GEMM and layernorm, which lower to highly optimized manual kernels). In aggregate, Inductor-generated kernels were 31\% faster than the handwritten kernels they replaced. \Figure{fig:inductor-subgraph} shows the breakdown of latencies of graph-mode sections, plotting the quotient of the latency with and without Inductor-generated kernels. Including eager-mode sections, the Inductor-optimized model achieved 4\% greater queries per second (QPS) improvement compared to baseline.

Fusion plays a major role in achieving the performance improvements. Analyzing the compilation result, the average fusion depth for Inductor-generated kernels is 3.9 operators. Notably, we observe data movement and type conversion operators are fused with element-wise ops, with the deepest fusion being 12 operators. Even when fusion depth is shallow, shape-specialized kernels from Inductor can often reach performance comparable to general manual kernels at runtime.

\subsection{Triton Production Footprint Ramp}
By enabling manual Triton kernel authoring and Inductor-Triton lowering, we increase the velocity for operator coverage improvement and performance
optimization. \Figure{fig:deploy-pg} plots the number of model types that Triton (both
manually authored and Inductor generated) has been rolled out to by Q1’25 and
Q2’25. Results from three product groups (PGs) are plotted. The number of
model types to which Triton was deployed tripled during this period across the three PGs.

\begin{figure}[t]
\centering
\includegraphics[width=\columnwidth]{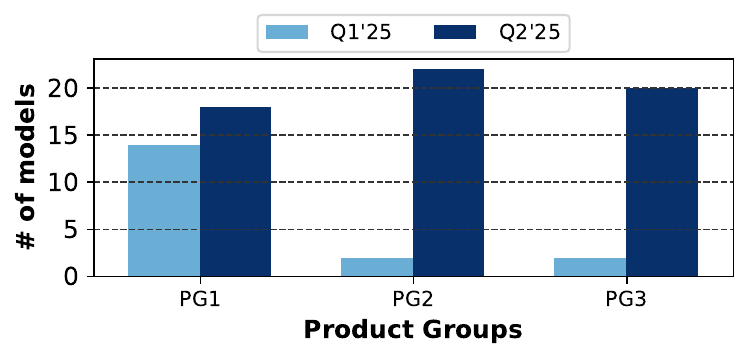}
\caption{Number of model types Triton deployed to on MTIA.}
\label{fig:deploy-pg}
\end{figure}

\Figure{fig:deploy-coverage} plots the percentage of layers and runtime
(GEMM excluded) that Triton accounts for in production, averaged
across models from \Figure{fig:deploy-pg}. Triton's production footprint
in terms of coverage of layers and execution time both roughly doubled.

\begin{figure}[t]
\centering
\includegraphics[width=.9\columnwidth]{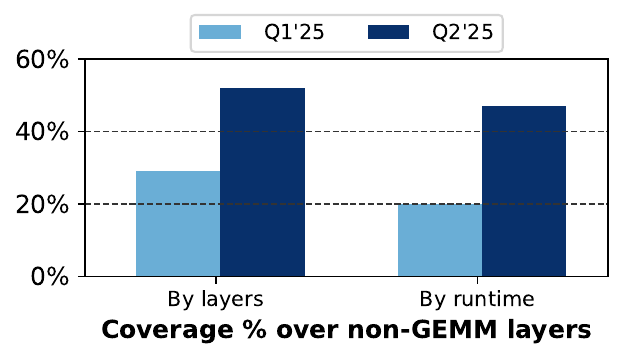}
\caption{Triton's production footprint in models on MTIA, measured by layer coverage and percentage of execution time (excluding GEMM).}
\label{fig:deploy-coverage}
\end{figure}

\section{EXPERIENCE}
In this section, we reflect on our experience in developing Triton for MTIA-2i and deploying Triton kernels in production at scale.

\subsection{Porting Triton Kernels from GPUs to MTIA}
Portability of Triton kernels between GPUs and MTIA was one of the main motivations for faithfully supporting the baseline Triton language. In a recent effort to enable production models on MTIA, we ported a suite of 16 Triton kernels from GPUs to MTIA. These kernels range widely in complexity, from tensor layout transformation to variants of fused attention.

Among the 16 kernels, 11 kernels are directly portable and provided good enough performance to unblock the use cases. A notable observation in this exercise was the reliance on Triton's undefined behavior on GPUs. In particular, \lstinline{tl.load} with non-block pointers takes an optional \lstinline{mask} parameter, which is often used to filter out out-of-bounds memory accesses. Users can provide an additional optional \lstinline{other} parameter, which is used to initialize the masked-out elements in the resulting tensor. Without specifying these \lstinline{other} values, these masked-out elements were initialized to zeroes on certain GPU backends, while we left them uninitialized on MTIA. This was the root cause for a variety of unexpected numeric errors. This same issue also led to out-of-bounds memory accesses, as indices were loaded from memory with uninitialized masked-out values. We fixed these kernels and verified the fixes were portable to GPUs. 

The remaining 5 kernels required various rewriting. They can be classified into 3 categories. 
\begin{itemize}
\item Nonportable pointer arithmetic operations. In two cases, the pointer arithmetic code could not be lowered to DMAs because it used modulo operations to wrap a tensor of pointers around the size of the tensor. This required only minor code changes, and the resulting kernels are portable to GPUs. 
\item Numeric differences between GPUs and MTIA. In one case, the numeric result of the kernel on MTIA is beyond the tolerance threshold. The root cause was in the differences on handling of mixed precision inputs between GPUs and MTIA. This issue requires explicit casting in the kernel, which is a minor code change.
\item Diverging optimization strategies. In cases where kernel complexity is high, relying solely on the compiler for performance optimization is infeasible. We rewrote the kernels to perform optimizations tailored for MTIA (e.g., explicit data reuse, different workload to PID mapping). 
\end{itemize}

As expected, we found that portability was one of the most significant benefits to using Triton on MTIA-2i accelerators. Using Triton made it possible to quickly support existing kernel libraries, which significantly accelerated our ability to support new models. Educating users to avoid using undefined behavior and using an MTIA-friendly coding style (e.g., to make sure the compiler can lower to DMAs when possible) also goes a long way in improving kernel portability across architectures.

Triton allowed us to trade a relatively small loss in peak performance for significant gains in kernel development velocity. For complex kernels, kernel-level optimization strategies can differ significantly, which justifies diverging kernels.

\subsection{Alternative DSLs}
Triton was not the only DSL we explored for MTIA architecture. In an earlier attempt, we developed a custom low-level DSL codenamed KNYFE~\cite{mtia2023} which predates Triton. KNYFE exposes MTIA's architectural details (e.g., CB, DMA, on-chip memory) directly to kernel authors, and enables performance that matches that of expert-tuned C++ kernels for GEMM and long-tail kernels. However, we found that the barrier to entry posed by the language's unfamiliar syntax was too high for all but the most experienced experts. This factor, combined with the need for portability and a higher level of abstraction, led us to abandon KNYFE in favor of Triton.

As the industry continues to drive the performance of ML models and kernels, recent DSL efforts have started to incorporate more architectural details into the language to regain some lost performance. Specifically, there are two approaches: (1) introducing architecture-tailored DSLs~\cite{cutedsl, gluon}, and (2) extending Triton or other languages with optional access to lower levels of abstraction~\cite{tlx}. Our Triton-MTIA extensions use the same strategy as TLX, but tailor the exposed operations and controls to the MTIA architecture. This composable approach preserves baseline Triton for general and portable kernels while allowing developers to selectively use MTIA-specific extensions in performance-critical code.

We also created a separate low-level DSL that directly exposes MTIA architectural features. Rather than maintaining an independent backend, this DSL reuses the same compiler infrastructure, including a subset of the dialects and passes developed for Triton-MTIA. More broadly, we are evolving the boundaries and interfaces between these dialects and passes so that future DSLs can reuse the appropriate compiler layers. This design should make it easier to support new programming models at different abstraction levels without duplicating the compiler stack.

\section{Conclusion}
The proliferation of specialized accelerator architectures, such as MTIA-2i, has transformed the landscape of large-scale machine learning deployments. While these custom accelerators offer important gains in performance and reductions in cost, they also introduce new programmability challenges, especially as operators adopt a heterogeneous mix of in-house and vendor-provided hardware. Our work demonstrates that, despite the diversity of programming models across accelerators, it is possible to bridge the gap between ML frameworks, kernels, and custom hardware by leveraging a flexible DSL like Triton. We hope our experience is useful for other accelerator efforts, and that it helps to build a stronger community around standardized DSLs to lower the barrier to adoption of new accelerators.

\bibliographystyle{IEEEtran}
\bibliography{paper}

\end{document}